# Impact of the interfacial Dzyaloshinskii-Moriya interaction on the band structure of one-dimensional artificial magnonic crystals: a micromagnetic study

R.Silvani[a,b], M. Kuepferling[a], S. Tacchi[c] and G. Carlotti[b]

a) *Istituto Nazionale di Ricerca Metrologica, Str. delle Cacce, 91, 10135 Torino, Italy*
b) *Dipartimento di Fisica e Geologia, Università di Perugia, I-06123 Perugia, Italy*
c) *Istituto Officina dei Materiali del CNR (CNR-IOM), Sede Secondaria di Perugia, c/o Dipartimento di Fisica e Geologia, Università di Perugia, I-06123 Perugia, Italy*

**Abstract**
We present the results of a systematic micromagnetic study of the effect of the Dzyaloshinskii-Moriya interaction (DMI) on the spin wave band structure of two one-dimensional magnonic crystals (MCs), both with the same periodicity *p=300 nm,* but different implementation of the DMI modulation. In the first system the artificial periodicity was achieved by modulating the interfacial DMI constant *D*, while in the second system also the sample morphology was modulated. Due to the folding property of the band structure in the dispersion relations of the magnonic crystals it is possible to extend the sensitivity of Brillouin light scattering towards weak DMI strength (*D* in the range from *0* to *0.5 mJ/m²*), by measuring the frequency splitting of folded modes in high-order artificial Brillouin zones, since the splitting increases almost linearly with the band index. For relatively large values of the DMI (*D* in the range from *1.0* to *2.0 mJ/m²*) the spin waves dispersion relations present flat modes for positive wavevectors, separated by forbidden frequency gaps whose amplitude depend on the value of *D*. These frequency gaps are more pronounced for the sample with morphology modulation. The non-reciprocal, localised, spatial profiles of these modes in both MCs are discussed with reference to spin waves in plain films and in isolated stripes of the same thickness.

1. **Introduction**

The study of the interfacial Dzyaloshinskii-Moriya interaction (DMI)[1,2] in magnetic thin films has become a very active research field in the last years. This interaction, at the origin of chiral magnetism, is an antisymmetric exchange coupling between two nearest spin electrons mediated by a heavy metal.[3,4,5] For future applications, an accurate measurement of its strength is indispensable, so that different methods have been developed to evaluate it, such as the study of the domain walls motion[6,7,8,9] and the detection of spin waves (SW) by Brillouin light scattering (BLS)[10,11,12,13,14,15,16,17,18,19] and all-electrical spin-wave spectroscopy (AESWS).[20,21,22,23] Several investigations have been published in the last years concerning plain films and multilayers,[24,25] while there are only a few investigations concerning patterned nanostructures and magnonic crystals.[26,27,28,29,30,31] The latter systems, where an artificial micrometric or submicrometric periodicity are usually realised by e-beam lithography, are attracting more and more interest in magnon spintronics for the realization of devices with unprecedented functionalities in information and communication technology.[32]

In a pioneering micromagnetic study, Ma and Zhou demonstrated the nonreciprocity of spin waves in nanostripe magnonic waveguides induced by the presence of interfacial DMI. They pointed out that the possibility of unidirectional spin wave propagation in a narrow frequency band could pave the way to a new class of compact nonreciprocal magnonic devices releasing the isolation and dynamic control of spin waves propagation.[26] Subsequently, M. Mruczkiewicz et al.[28] exploited both the frequency-domain method and micromagnetic simulations to investigate the impact of DMI on the ferromagnetic resonance (FMR) spectrum of spin wave modes in both isolated stripes and in one dimensional magnonic crystals (MCs) consisting of dense arrays of interacting stripes with a periodicity of 100 nm. They found that DMI caused the red-shift of some peaks in the simulated FMR spectrum, as well as the appearance of new peaks, because the presence of DMI modifies both the frequency and the spatial profiles of the modes. Another important contribution to the comprehension of the effect of interfacial DMI on spin waves was proposed more recently by R. A. Gallardo et al.[31] who analysed, through theoretical calculations and micromagnetic simulations, the evolution of the band structure of MCs realised by an artificial modulation of the DMI strength. They put in evidence the emergence of indirect band gaps and flat bands, discussing their dependence on both the DMI strength and the artificial periodicity. Very recently, Bouloussa et al.[33] presented a BLS investigation of magnonic modes in the presence of asymmetric dispersion induced by the interfacial DMI in a specially designed ultrathin CoFeB/Pt periodic structure. They showed that it was possible to detect experimentally magnonic modes of folded dispersion branches, although with a relatively low intensity if compared to the main (unfolded) branch.

In this work we aim at deepening the analysis of the effect of DMI on the band structure of one-dimensional MCs, considering two different implementations of the DMI and systematically varying the DMI effective constant $D$ in a wide range of values (between 0 to 2 $mJ/m^2$). The first sample consists of an array of wires realised modulating the DMI strength, while in the second one also the sample morphology, i.e. the thickness of adjacent wires, is modulated. Since we are interested in paving the way to direct measurements of real samples, the artificial periodicity is chosen to be $p=300$ nm, that is well fitted to the experimental investigation of two artificial Brillouin zones by BLS experiments. The band structure of the two MCs is calculated by numerical simulations whose results are discussed in details distinguishing two different regimes. In the weak-DMI regime, we point out that the presence of folded branches of the dispersion curves, induced by the artificial periodicity, might be exploited to access a larger wavevector range than in usual BLS experiments. This would enable an increased sensitivity of the BLS technique towards small values of the DMI strength, that are out of the reach for usual measurements in plain films. In the strong-DMI regime, instead, we analyse the non-reciprocal flat modes that appear in the dispersion curves, showing that more pronounced forbidden gaps are found for the MC where the sample morphology, and not only the DMI strength, is modulated. We also analyse the unusual spatial profiles of these peculiar magnonic modes, that are localised in specific regions of the MC, which, differently from ordinary modes with vanishing group velocity, present a non-reciprocal propagative character.

2. Methodology

As anticipated in the introduction, the current micromagnetic investigation involved two different one-dimensional MCs, whose characteristics are sketched in Fig.1: both samples have the same periodicity *p=300 nm*, but different implementation of the DMI modulation. In sample A (Fig. 1a) a thin ferromagnetic film (thickness *d=2 nm*) sits over a regular array of parallel heavy metal nanostripes of width *w=150 nm*. In sample B (Fig.1b), a continuous heavy metal substrate supports a ferromagnetic film with modulated thickness ($d_1=2$ *nm* and $d_2=4$ *nm*), forming an array of alternated parallel nanostripes of width *w=150 nm*. Note that the above geometrical parameters have been selected to match the characteristics of samples suitable for experimental investigation by BLS. In fact, the latter technique can access a range of wavenumbers up to about 20 rad/μm, that corresponds to two entire artificial Brillouin zones in the reciprocal space for our samples. The magnetic parameters of the ferromagnetic film are those typical of Permalloy, i.e saturation magnetization *Ms=730 kA/*m, exchange stiffness *A=10 pJ/m* and Gilbert damping *α=0.001*.

The dispersion relations of spin waves were calculated by the MuMax3 micromagnetic software,[34] systematically varying the amplitude of the interfacial DMI effective constant $D$ ($D_{ind}$ parameter in the MuMax3 code) from 0 to 2 $mJ/m^2$.

The simulated area of sample A (B) consisted of 2360×8×1(2) micromagnetic cells, corresponding to 40 periodicities. In both cases, the dimension of each cell was 5×5×2 nm³ and the periodic boundary conditions were applied in both in-plane directions, to account for the translational symmetry of the MCs. For sample A, $D$ was different from zero only for those cells of the magnetic film that are in contact with the heavy metal stripes, while for sample B, in order to account for a reduction of the interfacial DMI influence with thickness, the value of $D$ was reduced to a half in the upper part of the stripes (between 2 and 4 nm of thickness). In order to excite propagating spin waves in the simulated samples, they were first saturated along the y direction, i.e. parallel to the stripes, by an external field $B_{ext}=0.1\ T$ applied to the whole simulated region. Then, a sinc-shaped field pulse, $b(t) = b_0 \sin(2\pi f_0(t-t_0))/2\pi f_0(t-t_0)$ directed perpendicularly to the sample plane, with amplitude $b_0 = 10\ mT$ and maximum frequency $f_0 = 30\ GHz$, was applied in a region having a width of *40 nm* located in the centre of the simulated area (red region in Fig. 1a and 1b). The dispersion relations were finally obtained by recording the time evolution of the magnetization of each cell and then Fourier-transforming the out-of-plane component of the magnetization both in space and time. Please notice that the chosen geometry of excitation corresponds to the widely exploited Damon-Eshbach (DE) geometry, were the spin wave wavevector, directed along x, is perpendicular to the static magnetization that is directed along y.

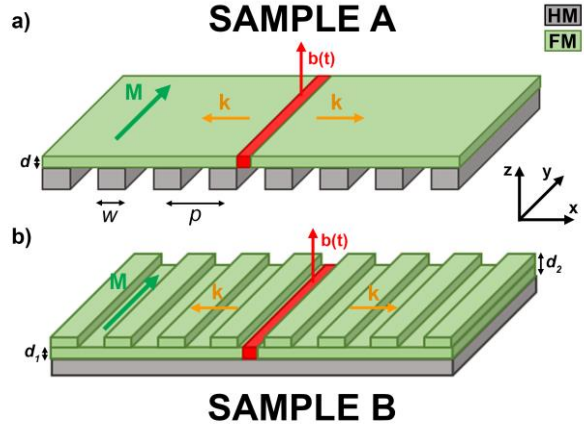

**Fig.1** Sketch of the two investigated one-dimensional magnonic crystals. Green and grey regions correspond to the ferromagnetic film and to the heavy metal substrate, respectively. In both cases the sample is magnetized along the y axis and spin waves, propagating along the ±*x* direction, are excited applying a perpendicular field pulse in a central line of width 40 nm (red region).

### 3. Dispersion curves of the plain film

As a first step towards the study of the above described MCs, we have analysed the effect of DMI on the propagation of spin waves the plain magnetic film, encompassing a wide range of $D$ values. It is well known from previous studies that DMI causes the presence of an extra term in the number of effective fields acting on the precessing magnetization, having the following expression: $\vec{H}_{DMI} = \frac{2D}{\mu_0 M_s}\left(\frac{\partial m_y}{\partial x}\hat{e}_x - \frac{\partial m_x}{\partial x}\hat{e}_y\right)$. Therefore, when one looks for spin waves, solving the linearized Landau-Lifshitz equation for ultrathin magnetic films in the DE propagation geometry, a simple analytical solution can be obtained in the following form:[35],[36]

$$\omega(k) = \omega_0(k) \pm \omega_{DMI}\ (k) =$$

$$\gamma\mu_0\sqrt{[(H_0 + Ak^2 + P(kd)M_s)(H_0 + Ak^2 + M_s - P(kd)M_s)]} \pm \frac{2\gamma}{M_s}Dk$$

(1)

Where $\gamma$ is the gyromagnetic ratio, $A$ the exchange stiffness constant and $M_s$ the saturation magnetization. The plus or minus sign applies to the case of the external field $H_0$ directed either parallel or antiparallel to the y axis. Moreover, the dipolar term $P(kd) = 1 - \frac{1-e^{-|kd|}}{|kd|}$ in the case of ultrathin films, where $kd<<1$, reduces to $P(kd) = \frac{|kd|}{2}$ as a result of the series expansion of the exponential term.

As seen in Fig. 2a, for $D=0$ the simulated dispersion relation of a $d=2$ nm thick film is perfectly symmetric for positive or negative k-vector. When $D$ increases, however, spin waves with positive (negative) $k$ considerable decrease (increase) their frequencies. In Fig. 2b one can see that for a film of thickness $d=2$ nm the results obtained by our micromagnetic simulations (purple points) agree very well with those calculated by the analytic expression of Eq (1), for either $D=0$ or *1.6 mJ/m²*. If one then considers in the micromagnetic simulations a film divided in two layers of equal thicknesses $d_1=d_2=2$ nm with different values of $D$, namely *1.6 mJ/m²* and *0.8 mJ/m²* (brown points), the agreement with the analytic results is achieved assuming in Eq. (1) a film of thickness $d_1+d_2=4$ nm and *D=1.2 mJ/m²*, corresponding to the average of its two values in the sublayers. The above results will be exploited further on to shed light on the behaviour of spin waves in the two MCs.

From the above results it turns out that if one wants to measure $D$ in a magnetic film from the frequency asymmetry of counter-propagating spin waves, it is convenient to achieve the largest accessible value of $k$. However, in a BLS experiment, where this is achieved by simply measuring the different frequency of the Stokes and anti-Stokes peaks, one has to face the limitation of a maximum accessible spin wave k-vector of about 20 *rad/μm*. Therefore, when the DMI amplitude $D$ is relatively small (for instance below about 0.2 *mJ/m²*) the frequency splitting may be lower than 0.1-0.2 *GHz*. i.e. difficult or impossible to detect experimentally, taking into account the limited frequency resolution of the apparatus and the finite linewidth of magnon peaks (that usually depends also on the structural and morphological quality of the interfaces and of the whole sample).

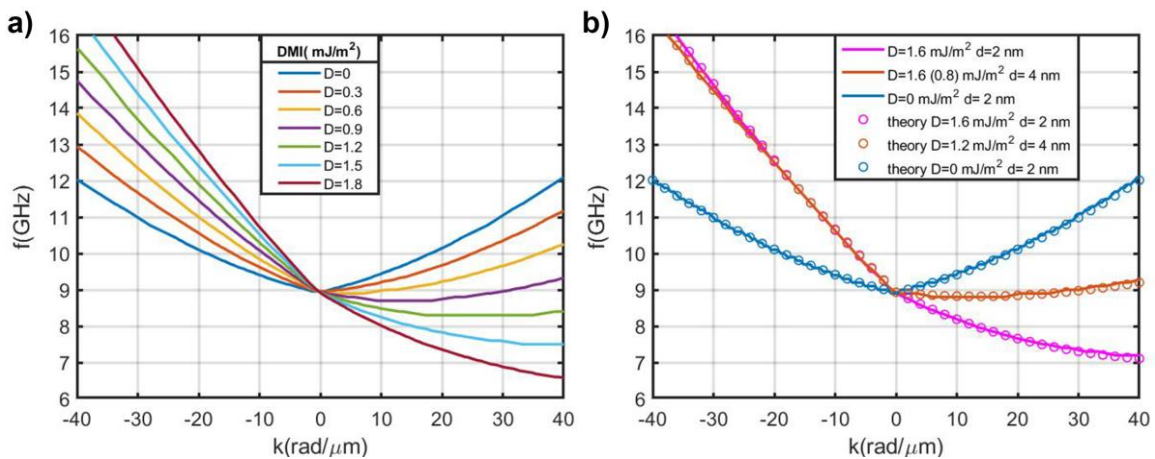

**Fig.2** a) Micromagnetic simulations of spin wave dispersion relations in a plain film, *2 nm* thick, for different values of the effective DMI constant $D$. The curves become strongly non-reciprocal with the increasing $D$. In b) the results of micromagnetic simulations (continuous line) are compared with the analytical expression of the spin waves dispersion relation derived from Eq. 1 (points).

## 4. Band structure of the magnonic crystals

The effect of the presence of a sizeable DMI on the dispersion relations of the two MCs considered in this paper can be seen in Fig. 3, where the results of the micromagnetic simulations are shown in grey-intensity scale for different values of *D*, in a wavevector range spanning over three artificial Brillouin zones. In absence of any DMI (*D=0*, left panels in Fig. 3) the typical dispersion curve of the Damon-Eshbach mode in a plain ferromagnetic film is seen for sample A, while for sample B the formation of magnonic bands, due to the morphological periodicity, is observed. When the value of *D* is lifted from zero to about *0.5-0.6 mJ/m²* the periodicity of the DMI strength manifests itself in the appearance of magnonic bands also in sample A, but the forbidden band gaps are very small in both samples. In addition, the curve with maximum intensity, that corresponds to the dispersion curve of the plain film, now exhibits a clear non-reciprocity, having different slopes for negative or positive wavevectors *k*. The situation becomes qualitatively different when the value of *D* increases above about *1 mJ/m²* (Fig.3 right panels): a larger number of high-frequency bands, characterized by sizeable band gaps, become visible. Moreover, some of these bands at low frequencies, more intense for positive *k*, are flat, reflecting the presence of localised spin wave modes at fixed frequencies, i.e. with zero group velocity, separated by ample forbidden band gaps. Let us now, in the next two sections, discuss in details the characteristics of the magnonic bands in each of the two mentioned regimes, i.e. those of relatively small and relatively large DMI.

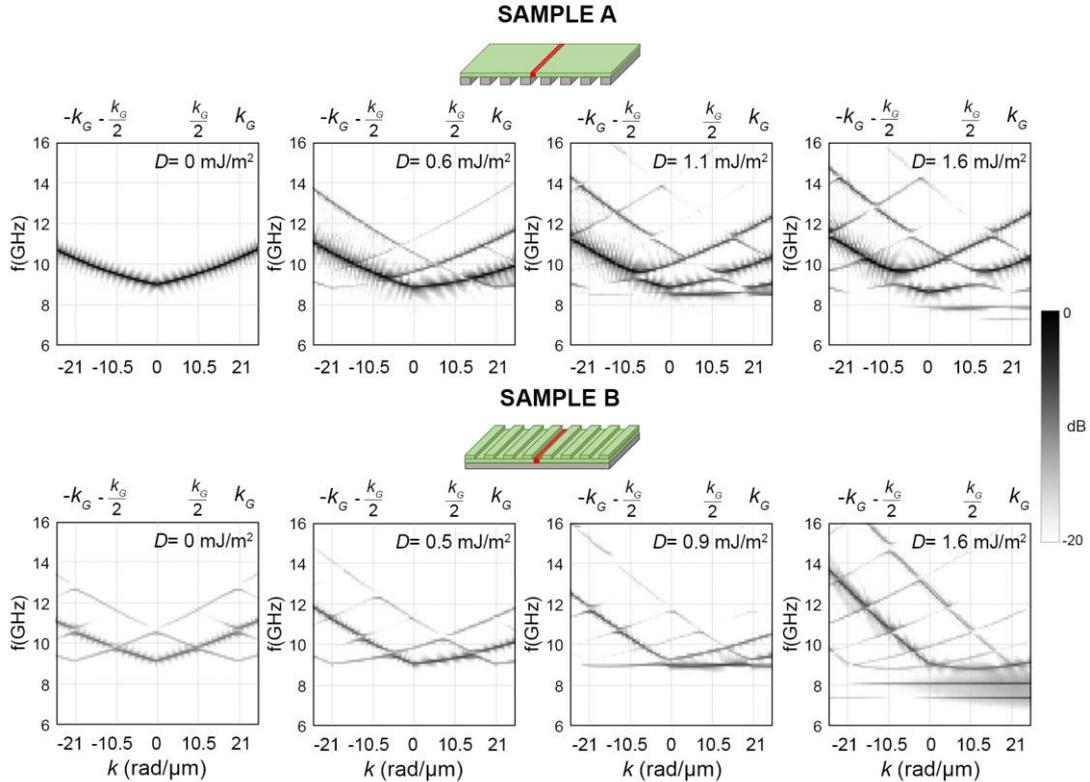

**Fig.3** Dispersion relations of the two analysed MCs, i.e sample A (top panels) and sample B (bottom panels), obtained by micromagnetic simulations, for different value of the interfacial DMI constant *D*.

## 4.1 Region of small values of the effective DMI constant (D <0.5 mJ/m²): increase of the spin wave frequency asymmetry in high-order artificial Brillouin zones

As anticipated in the previous section, in MCs with relatively small values of the DMI constant, the most intense curve in the dispersion relations is similar to that of the corresponding plain film and the tiny frequency asymmetry $\Delta f$ for spin waves with $\pm k$, i.e the frequency difference between the Stokes and the anti-Stokes peaks in BLS spectra, can be difficult or even impossible to be measured. This is illustrated in Fig 4 for sample A, assuming $D=0.1$ $mJ/m^2$. The main dispersion curve is plotted in green and the white area indicates the region accessible to BLS. One can see, for instance, that considering spin waves in the middle of the first Brillouin zone, with a wavevector $k_0= 5.2$ $rad/\mu m$ (red arrow in Fig. 4, corresponding to an angle of incidence $\theta=13°$ in a BLS experiment) the corresponding frequency asymmetry $\Delta f$ would be negligible, i.e. well below the sensitivity of BLS of about 0.1-0.2 GHz. The situation is not much different if one choses a larger angle of incidence in BLS experiments, because the maximum achievable value of $k_0$, corresponding to an angle of incidence $\theta=70°$, is anyway limited to about $20$ $rad/\mu m$ and the $\Delta f$ would still be too small. However, it is worth to notice that if one would be able to access higher order Brillouin zones (shadowed areas in Fig. 4), then the frequency asymmetry would appreciably increase above the threshold required for BLS experiments. Quantitatively, this problem can be analysed in details in Fig. 5, left panel, where we have collected the values of the frequency asymmetry observed in the first five Brillouin zones, for different values of $D$ in the range between $0.1$ and $0.3$ $mJ/m^2$. One can see that the frequency asymmetry $\Delta f$ increases almost linearly with the magnonic band index $n$ so that even for $D$ as low as $0.1$ $mJ/m^2$ a substantial splitting, easily measurable by BLS, could be achieved for the modes with index $n=3$ or $4$. To this respect, although direct access to the high index BZs is out of range of accessible wavevectors in a BLS experiment (shadowed regions in Fig.4), one may consider to measure the folded branches that occur in the accessible region (blue curves within the white area of Fig. 4) thanks to the reciprocal-space periodicity that originated from the modulation of the MC in the real space. This means that the wavevector conservation in a BLS experiment would be satisfied not only by spin waves with wavevector $\pm k_0$, but also by spin waves with values of the wavevector $\pm k_n$ that can be obtained summing or subtracting to $\pm k_0$ an integer number of grating wavevectors $k_G=2\pi/p$. This is illustrated on the right side of Fig. 4, were the amplitude of the wavevector $k_n$ corresponding to each magnonic band is indicated. We have then estimated the amplitude of the frequency asymmetry $\Delta f_n$ of each magnonic band, using Eq. 1, valid for a plain film, and inserting the values of $k_n$ for each magnonic band, as well as the values of the DMI constant, averaged over the volume of the film of thickness 2 and 4 nm for sample A and B, respectively:[37]

$$\Delta f_n = \frac{2\omega_{DMI}}{2\pi} = \frac{2\gamma}{\pi M_s} <D> k_n \qquad (2)$$

The result of this approximation, reported as a continuous line in Fig. 5a, indicates that this is in good agreement with the results of micromagnetic simulations for sample A. However, it is important to point out that the access to the high order folded branches might be challenging by conventional BLS, because the expected cross section of high-order modes decreases with the mode index $n$, so that long acquisition time and very good signal to noise ratio would be necessary. Alternatively, one may overcome this difficulty using micro-focused BLS and integrating a properly designed transduction antenna over the sample surface, to excite spin waves at the frequency of the folded branches by a microwave signal generator. In such a way, the sensitivity of BLS could be in principle increased towards values of $D$ as small as *0.1 mJ/m²*, that are out of the reach for BLS measurements in plain films. Finally, it is interesting to notice that for sample B the frequency asymmetry $\Delta f_n$ would be even larger than in sample A. In fact, the larger slope of the data of Fig.5b for sample B reflects the fact that in this sample the DMI is present in all its volume, rather than only in alternated regions, so that the non-reciprocity of the dispersion curves is more pronounced than for sample A. However, one can see that as $D$ is lifted above *0.2 mJ/m²*, the micromagnetic simulations (points) indicate that the dependence of $\Delta f_n$ on the magnonic index $n$ for sample B is not monotonous and the approximation based on Eq. 2 is not valid anymore. This reflects the complex structure of the dispersion curves and the presence of magnonic gaps, whose amplitude is enhanced in this sample by the modulation of the surface morphology, as anticipated in the introduction.

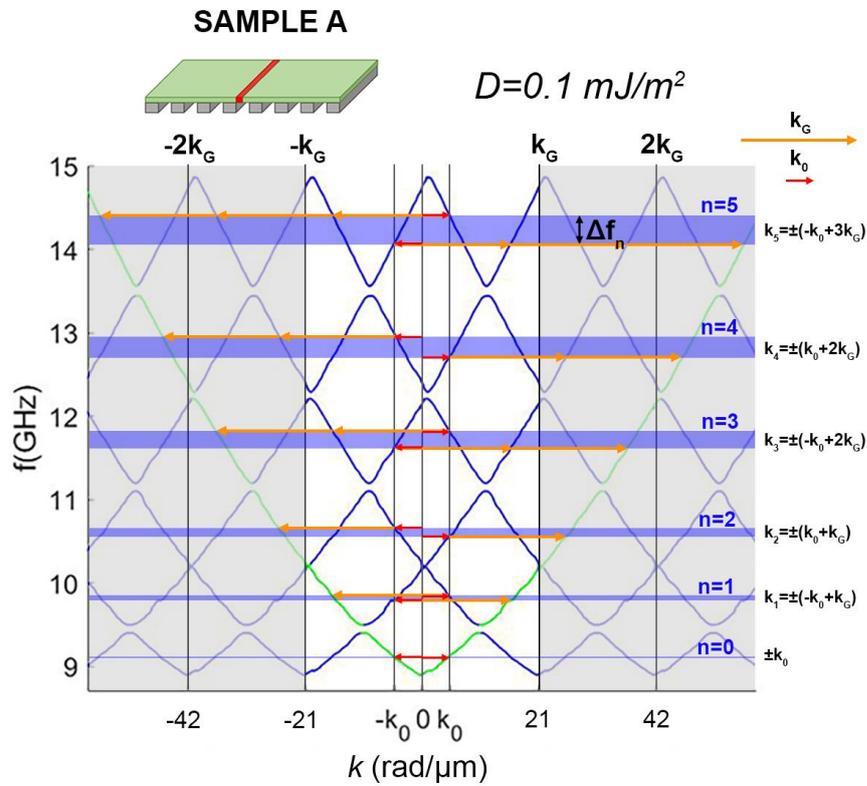

**Fig.4** Magnonic band structure of sample A obtained by micromagnetic simulations for $D=0.1$ *mJ/m²*. The green curve is the *principal* dispersion relation, while the blue curves are the periodic dispersion relations due to the presence of the artificial periodicity. Shadowed areas are those out of the reach of a BLS experiment. If one considers to detect spin waves with wavevector $\pm k_0$ (red arrows) within the first Brillouin zone, it is seen that the corresponding frequency asymmetry is negligible. However, taking into account the periodicity of the magnonic crystal and the possibility of adding an integer number of grating vector $\pm k_G$ (orange arrows), it is possible to access spin waves in higher-order Brillouin zones, characterized by larger values of the wavevectors ($\pm k_n$) and larger frequency asymmetry $\Delta f_n$

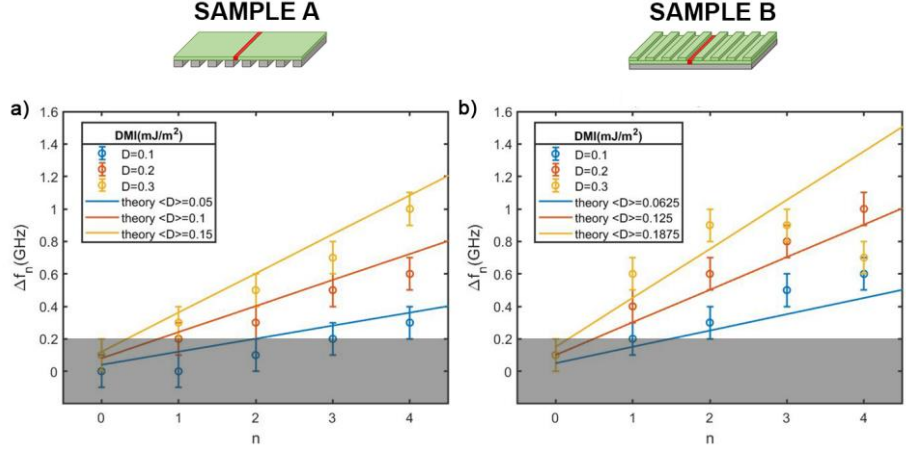

**Fig 5** Frequency asymmetry of spin waves with $\pm k_n$ for relatively small value of *D* in sample A (a) and in sample B (b). The points are results of the micromagnetic simulations, while the lines are obtained from the theoretical formula valid for a plain film, assuming an average value of *D*, as explained in [37].

### 4.2 Region of large values of the effective DMI constant (D >1 mJ/m$^2$): appearance of non-reciprocal flat modes

In the region of relatively strong DMI (*D* larger than about *1 mJ/m$^2$*), there is an important qualitative change in the dispersion curves of the MCs shown in Fig. 4: quasi-flat modes, separated by substantial forbidden band gaps, appear at low frequency, more intense for positive values of k, reflecting the existence of spin waves localised in specific regions of the MCs. Their frequency is fixed because their group velocity is zero and they are similar to stationary modes confined in isolated magnetic stripes. In fact, the presence of the artificial modulation, is such that the effective field acting on the precessing spins is remarkably different in adjacent stripes. Interestingly, it is seen that in sample B the above characteristics are even more evident and quantitatively relevant than in sample A, as shown in Fig. 6, where we plot the dependence of the band gap amplitude between the two low-frequency flat modes on the strength of the DMI constant *D*, ranging from 1 to 2 *mJ/m$^2$*. One can see that the gap amplitude takes values in the range between 0.3-0.7 *GHz* and even if the periodic modulation of the DMI is more pronounced in sample A than in sample B, the presence of a surface morphology in the latter sample enhances the gap amplitude. Another interesting characteristic, evident in Fig. 6, is that for both samples the gap amplitude keeps constant in a wide range of *D* values, reflecting the simultaneous decrease in frequency of both flat modes when *D* increases. This is similar to what is found in isolated wires in presence of DMI, where the different stationary modes decrease in frequency with almost the same slope as a function of *D*.[38] This means that a direct BLS measurement of the frequency separation of the two low-frequency flat modes, does not provide direct access to a quantification of *D*. For this purpose, it is still necessary to measure the frequency asymmetry between the dispersive spin waves at higher frequencies.

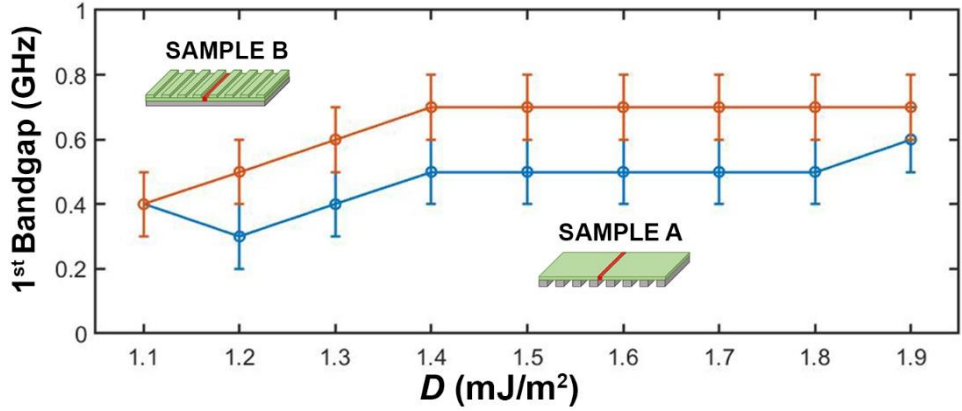

**Fig 6** Amplitude of forbidden band gap between the two low-frequency flat modes for sample A and sample B for relatively large values of the DMI constant *D*.

To better understand the characteristics of the flat modes and their spatial profile, we have analysed the dynamics of the two MCs when an external sinusoidal signal is applied to excite the system at the frequency of the low-lying flat mode (7.3 GHz for sample A and 7.4 GHz for sample B). The obtained time evolution, in an oscillation period *T*, of the spatial profile of the $m_z$-component of the dynamic magnetization, for a value of D=*1.6 mJ/m²*, is shown in Figs. 7 and 8. One can see that for sample A (Fig. 7) the mode is essentially localized in the regions of the magnetic film sitting over the heavy metal wires. Moreover, the time evolution of this stationary mode inside each stripe is such that its spatial profile is characterized by a couple of nodes and antinodes, corresponding to an effective wavelength of about half width of the stripes. Remarkably, the shape of the profile shifts along the positive *x*-direction (green arrows in Fig. 7) as a function of time, for both stripes on the right and on the left of the excitation region. This can be understood recalling Fig.2b, where the dispersion relation in a thin film with and without DMI is simulated. One can see that spin waves at 7.3 GHz exists only in the positive *k* region relative to the film with *D=1.6 mJ/m²* (violet curve). This explains why the flat mode profile propagates to the right and is localised only in the part of the MC where the DMI is present, i.e. where the film is supported by the heavy metal wires. In the other regions (white areas in Fig. 7) the amplitude is very low, without oscillations, so the spatial profile is similar to that of the electron wavefunction in a potential barrier, where it rapidly decays (this is at the origin of the well-known tunnel effect). In any case, the fact that the mode cannot propagate through the whole crystal is consistent with a vanishing group velocity and a flat dispersion curve. A similar situation is also found in sample B, as illustrated in Fig. 8. The flat mode at 7.4 GHz is localised in the thinner regions of the sample (shadowed areas) and again this can be understood looking at Fig. 2b where we show the simulated dispersion relations of two films with *2 nm* and *4 nm* thickness: spin waves at 7.4 GHz exist only in the thinnest film and only for positive wavevectors. Finally, from a comparison of the profiles of Figs. 7 and 8 one may notice that in sample B the pseudo-stationary mode has a larger effective wavelength than in sample A, although the width of the localization region is the same (150 nm) in both samples. This is a consequence of the different mechanism of localization and the different boundary conditions felt by the precessing spins at the edges of the localization regions.

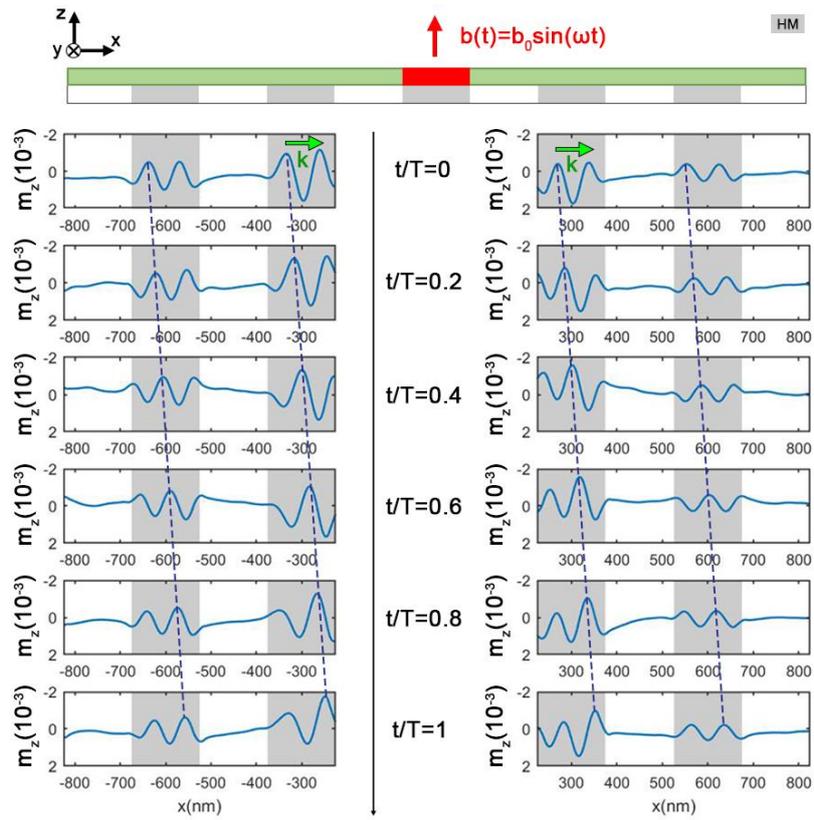

**Fig.7** Spatial profile of the dynamical magnetization component perpendicular to the film plane, for the flat mode of sample A (7.3 GHz). The shadowed areas, where the mode amplitude is localised, correspond to the regions of the ferromagnetic film sitting over the heavy metal wires. The vertical (dashed) lines are guides to the eye and indicate that the spatial profile of the mode propagates towards the right on both sides of the excitation region (red area in the top inset). Instead, in the white areas where the DMI is not present, the mode amplitude is very small (resembling the well-known "tunnel effect" of electrons through a potential barrier).

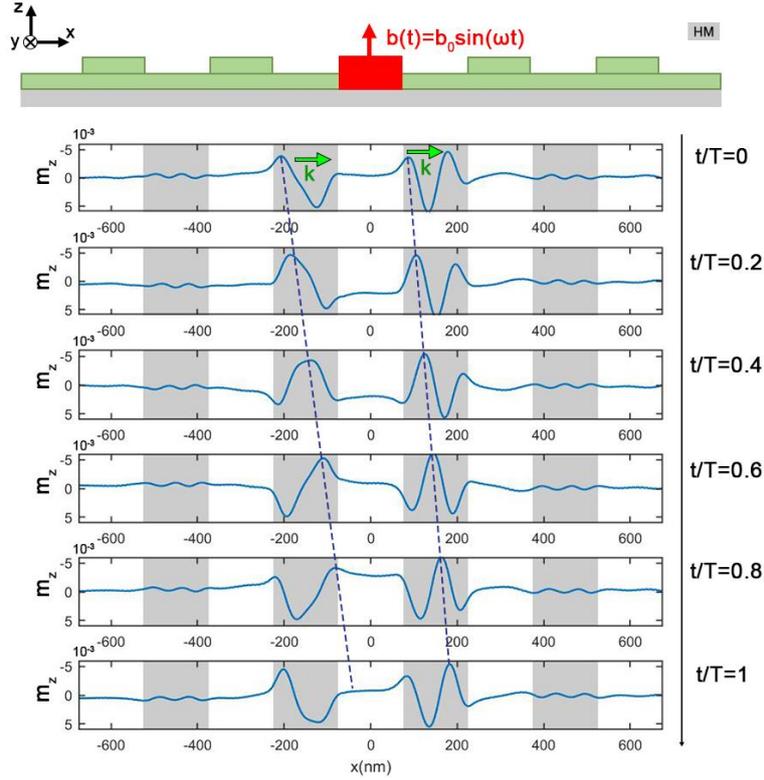

**Fig.8** Spatial profile of the dynamical magnetization component perpendicular to the film plane, for the flat mode of sample B (7.4 GHz). The shadowed areas, where the mode amplitude is localised, correspond to the thinner regions of the sample. The vertical (dashed) lines are guides to the eye and indicate that the spatial profile of the mode propagates towards the right on both sides of the excitation region (red area in the top inset). Instead, in the white areas, i.e. the thicker regions of the sample characterized by a reduced value of the DMI, the mode amplitude is very small (resembling the well-known "tunnel effect" of electrons through a potential barrier).

## 5. Conclusions

In this work we have systematically studied the effect of the interfacial Dzyaloshinskii-Moriya interaction on the dispersion relations of spin waves in two different one-dimensional magnonic crystals having a periodicity tailored to possible experimental investigation by BLS. In one system the implementation of the artificial periodicity was based on the modulation of the DMI strength, while in the second one also the sample morphology was modulated. The micromagnetic simulations, carried on using the software MuMax3, showed in both samples two different regimes, corresponding to either small or large values of the DMI strength, expressed through the value of the constant $D$.

For values of $D$ in the range from *0* to *0.5 mJ/m²*, we suggest to take the opportunity of the folding property of the dispersion relations of the magnonic crystals to extend the BLS sensitivity towards weak DMI strength, by measuring the frequency non-reciprocity of folded modes in high-order artificial Brillouin zones, since the splitting increases almost linearly with the band index.

For relatively large values of the DMI ($D$ in the range from *1.0* to *2.0 mJ/m²*) the spin waves dispersion relations present flat modes for positive wavevectors, separated by forbidden frequency gaps whose amplitude depends on the value of *D*. Following the time evolution of the dynamic magnetization of these modes, one realizes that they are confined in specific regions of the samples and, within such regions, propagate only along the positive *x* direction. These characteristics have been explained looking at the spin wave dispersion relations of ferromagnetic films with the same thickness and magnetic parameters of the different portions of the magnonic crystals. Finally, we found that the above characteristics and the band gap amplitude are more marked in the magnonic crystal realised by a surface modulation, changing the thickness of the adjacent ferromagnetic stripes (sample B), rather than in the one where there is only a modulation of the DMI strength in adjacent stripes (sample A). We are confident that these results will stimulate the realization of BLS experiments on magnonic crystals characterized by either relatively-weak or relatively-strong DMI, that are still lacking in the literature.

## 6. Acknowledgements

Financial support from the EMPIR programme 17FUN08-TOPS, co-financed by the Participating States and from the European Union's Horizon 2020 research and innovation program, is kindly acknowledged.